\documentclass{article}

\usepackage{amsmath,amssymb}
\usepackage{graphicx}

\newcommand{\newc}{\newcommand}
\newc{\gev}{\,GeV}
\newc{\sgn}{\mr{sgn}\,}
\newc{\ra}{\rightarrow}
\newc{\rpv}{$\mathrm{\not\!R_p}$}
\newc{\met}{$\not\!\!E_T$}
\newc{\rp}{$\mathrm{R_p}$}
\newc{\real}{\mathcal{R}e}
\newc{\alsm}{{\displaystyle \sum_{\alpha=1,2}}}
\newc{\besm}{{\displaystyle \sum_{\beta=1,2}}}
\newc{\al}{\alpha}
\newc{\ga}{\gamma}
\newc{\de}{\delta}
\newc{\cw}{\cos\theta_w}
\newc{\ssw}{\sin^2\theta_w}
\newc{\ccw}{\cos^2\theta_w}
\newc{\cbe}{\cos\beta}
\newc{\sbe}{\sin\beta}
\newc{\sh}{\hat{s}}
\newc{\sa}{\sin\al}
\newc{\ca}{\cos\al}
\newc{\bv}{$\mathrm{\not\!B}$}
\newc{\lv}{$\mathrm{\not\!L}$}
\newc{\ie}{{\it i.e.\/}\ }
\newc{\lam}{\lambda}
\newc{\cht}{\tilde{\chi}}
\newc{\upt}{\tilde{u}}
\newc{\elt}{\tilde{\ell}}
\newc{\hgt}{\tilde{H}}
\newc{\nut}{\tilde{\nu}}
\newc{\dnt}{\tilde{d}}
\newc{\psb}{\bar{\psi}}
\newc{\rtt}{\sqrt{2}}
\newc{\mut}{\tilde{\mu}}
\newc{\mr}{\mathrm}
\newc{\bath}{\bar{\theta}}
\newc{\tht}{\theta}
\newc{\JC}{{\bf J}}
\newc{\lra}{\longrightarrow}
\newc{\eg}{{\it e.g.\,}}
\newc{\barr}{\begin{eqnarray}}
\newc{\earr}{\end{eqnarray}}
\newc{\beq}{\begin{equation}}
\newc{\eeq}{\end{equation}}
\newc{\me}{\mathcal{M}}
\newc{\dbm}{\partial_\mu}
\newc{\sgm}{\sigma_\mu}

\def\barr{\begin{array}}
\def\earr{\end{array}}
\def\be{\begin{equation}}
\def\ee{\end{equation}}
\def\ra{\rightarrow}

\catcode`@=11 % This allows us to modify PLAIN macros.
\def \gsim{\mathrel{\mathpalette\@versim>}}
\def \lsim{\mathrel{\mathpalette\@versim<}}
\def \@versim#1#2{\lower0.4ex\vbox{\baselineskip\z@skip\lineskip\z@skip
     \lineskiplimit\z@\ialign{$\m@th#1\hfil##\hfil$%
     \crcr#2\crcr\sim\crcr}}}
\catcode`@=12 % at signs are no longer letters
\def\gev{\: \rm GeV}
\begin{document}
\setcounter{page}{0}
\renewcommand{\thefootnote}{\fnsymbol{footnote}}
\thispagestyle{empty}

\begin{titlepage}
%\vspace{-2cm}
%\begin{flushright}
%MRI--P--020704\\[2ex]
%{\large \tt hep-ph/yymmnnn}\\
%\end{flushright}
%\vspace{+2cm}

\begin{center}
 {\Large{\bf Thermodynamics of rotating black hole in the presence of cold dark matter}}\\
\vskip 0.6 cm
{\bf Swapan Kumar Majhi }
        \vskip 1cm
{\it Ranaghat College, Nadia, West Bengal-741201, India}
        \vskip 0.5cm
{E-mail: majhi.majhi@gmail.com}
\\
\vskip 0.2 cm
\end{center}
\setcounter{footnote}{0}
\begin{abstract}\noindent
	We have computed the thermodynamic properties of a rotating black holes in the presence 
	of cold dark matter. The dependence of temperaure, Gibbs free energy, specific heat on 
	the horizon radius have been studied for various values of critical density ($\rho_c$) 
	of the cold dark matter. 
	The studies shows that only small values of critical density of cold the dark matter, 
	black hole are stable.
\end{abstract}
\end{titlepage}

\setcounter{footnote}{0}
\renewcommand{\thefootnote}{\arabic{footnote}}

\setcounter{page}{1}
\pagestyle{plain}
\advance \parskip by 10pt

\section{Introduction}
%%%%%%%%%%%%%%%%%%%%%%%%
 In recent years, the study of dark matter(DM) is an emerging area of research
which basically connects cosmology and particle physics \cite{dm-china,zhou}.
Current observations indicate
that DM is nearly invisible and hence can not be directly observed but its existence can be
observed through indirect evidence \cite{jhoort}, for example
microwave background radiation meassurements \cite{zshen}
weak gravitational lensing effects \cite{jzhang,nnweinberg},
measurements of galaxy rotation \cite{rubin,ekaterina,ms.roberts},
large scale structure measurements \cite{lfang,lgao},
galaxy light-to-mass ratios \cite{zzhong}
and cluster dynamics measurements \cite{tatagliaferro,fzwicky}.
Among various dark matter models, the cold dark matter(CDM) model is one of the
predominant models for dark matter \cite{SCDM_nfw,UniDenProf}.
It is composed of neutral weakly interacting heavy particles \cite{xbi} and
the number of DM is very large \cite{parade}.
Vera Rubin's \cite{rubin} observations of the motion of stars within galaxies is one of the
prime evidence in early 1980s and it predicted more concentrated matter distribution
in the galaxies than predicted by gravitational effects which points the strong support for the
theory of dark matter. The Wilkinson Microwave Anisotropy Probe (WMAP) produced the first image
of the infant universe and accurately measured the cosmological parameters in 2003.
The same result obtained by the Sloan Digital Sky Survey (SDSS) which confirm the existence of
dark matter \cite{cseife}.
The rapid developement of gravitational wave astronomy\cite{yhu, zzhu,jliu,bpabbott1,bpabbott2,bpabbott3} and the successful 
capture of black hole images in 2019\cite{kakiyama1,kakiyama2,kakiyama3}, black holes have become another favourite 
topic of research area among physicists. It has been noticed \cite{dlynden,aeckart,kgebhart,jkormendy} that 
there exist supermassive black holes along with the dark matter particles in the 
central regions of galaxies. Both black hole and dark matter can form stable system in the 
central region of galaxies\cite{jkormendy}. The detailed study of their interactions and 
dynamics will reveal the nature of dark matter as well as black holes.

One of the interesting feature of black hole physics is their thermodynamic properties. 
The pioneer works of Hawking and Bekenstien \cite{bekstn}-\cite{bar-car-hwk} opened up 
the new direction in this context. First time, they have pointed out that
there is an interesting resemblance between the spacetimes with horizons and the
thermodynamical systems with well defined temperature and entropy. In these works, they have
showed that the black hole behaves like a thermal systems and laws of black hole mechanics are
basically same as laws of thermodynamics.
Very first time, Hawking {\it et. al.} \cite{bekstn}-\cite{bar-car-hwk} tried to construct entropy
of the black hole event horizon using phenomenological models. Black hole entropy and thermodynamic
properties have been studied extensively various cases like Schwarzschild space time \cite{bekstn}-\cite{wald},
Reissner-Nordstrom black hole \cite{good_ong} and rotating and non-rotating black hole \cite{fatima}-\cite{fjaved}.
The study of thermodynamic properties of black hole has revealed  many aspects of their physics and
hence it is very important to investigate the black holes presence of dark matter and and their
thermodynamical properties.

In this article, we begin with a short review of spacetime metric in section \ref{space-time-metric}.
In section \ref{thermodyn-computation}, we have calculate various thermodynamic variables like 
Bekenstein-Hawking entropy, Hawking temperature, Gibbs free energy and specific heat etc. 
with numerical plots and their discussions. At the end, we summerise. We have used units which fix
the speed of light and gravitational constant $G = C = 1$ and also used the scale radius
$R_s = 1$ for simplicity, through out our numerical analysis.

%%%%%%%%%%%%%%
{\section{Space-time metric of rotating black hole in the presence of Dark Matter}
\label{space-time-metric}}
%%%%%%%%%%%%%%
The distribution of dark matter halos within galaxies can be described using several parameters, 
including scale radius and critical density, which emerge from the interactions among dark matter 
particles.
When a black hole resides at the geometric center of a dark matter halo, there can be series of 
interactions between the black hole and the encompassing dark matter halo.
The dark matter-black hole systems can be precisely characterized by employing 
the steady-state approximate black hole metric.
In this work, we consider the cold dark matter halos as determined by the
cold dark matter model.

According to Navarro-Frenk-White (NFW) profile, the density distribution  of the dark matter halo
in the cold dark model is given by \cite{SCDM_nfw, SCDMH,konoplya}
\begin{equation}
	\rho_{NFW} = {\rho_c \over {r\over R_{s}} (1+{r\over R_{s}})^2}
\label{NFW_profile}
\end{equation}
 where $\rho_c$ denotes the critical density of the dark matter halo and $R_{s}$ represents
 the scale radius of the dark matter halo. In a dark matter halo scenario, the spacetime 
 metric of a rotating black hole \cite{ref_cdm_metric} is 
\begin{eqnarray}
	\label{cdm_metric}
	ds^2 &=& -\Bigg[1- {r^2 + 2 M r - r^2 \Big(1+{r\over R_{s}}\Big)^{-{8\pi \rho_c R^3_{s}\over r}}\over \Sigma^2} \Bigg] dt^2 \nonumber \\[1ex]
	&& + {2 a \sin^2\theta \over \Sigma^2} \Bigg[r^2 + 2 M r - r^2 \Big(1+{r\over R_{s}}\Big)^{-{8\pi \rho_c R^3_{s}\over r}} \Bigg] dt d\phi \nonumber \\[1ex]
	&& + {\sin^2\theta \over \Sigma^2} \Big[(r^2+a^2)^2 - \Delta\,a^2\sin^2\theta \Big] d\phi^2 + {\Sigma^2 \over \Delta} dr^2 + \Sigma^2 d\theta^2
\end{eqnarray}
with 
\begin{eqnarray}
	\Sigma^2(r,\theta) &=& r^2 + a^2 \cos^2\theta \\
	\Delta(r) &=& a^2 - 2 M r + r^2 \Big(1+{r\over R_{s}}\Big)^{-{8\pi \rho_{c} R^3_{s}\over r}}
	\label{delta_cdm}
\end{eqnarray}
where $M$ is the black hole mass, $a$ is rotational parameter. As the dark matter critical 
density ($\rho_c$) goes to zero, the above metric reduces to $Kerr's$ rotational black hole meric. 

We have calculated the scalar invariants like Ricci scalar and Kretschmann scalar. The Ricci scalar
is proportional to $\Sigma^{-2}$ and has a simple form
\begin{equation}
        R = {-2 +\partial^2_r \Delta(r) \over \Sigma^2(r,\theta)}
\end{equation}
 whereas the Kretschmann scalar $R_{\mu\nu\rho\sigma} R^{\mu\nu\rho\sigma}$ is proportional
 to $\Sigma^{-6}$ i.e.
 \begin{equation}
         R_{\mu\nu\rho\sigma} R^{\mu\nu\rho\sigma} = {f(r,a,\theta) \over \Sigma^6(r,\theta)}.
\end{equation}
The function $f(r,a,\theta)$ is regular as $\theta = {\pi \over 2}$ and $r \rightarrow 0$ and  
too large to reproduce here. At $\theta = {\pi \over 2}$ and $r \rightarrow 0$, 
metric eqn.(\ref{cdm_metric}) possesses a singularity called ring singularity which also 
presents in Ricci scalar and Kretschmann scalar.

%%%%%%%%%%
\begin{figure}
%\centering
\includegraphics[width=7.5cm]{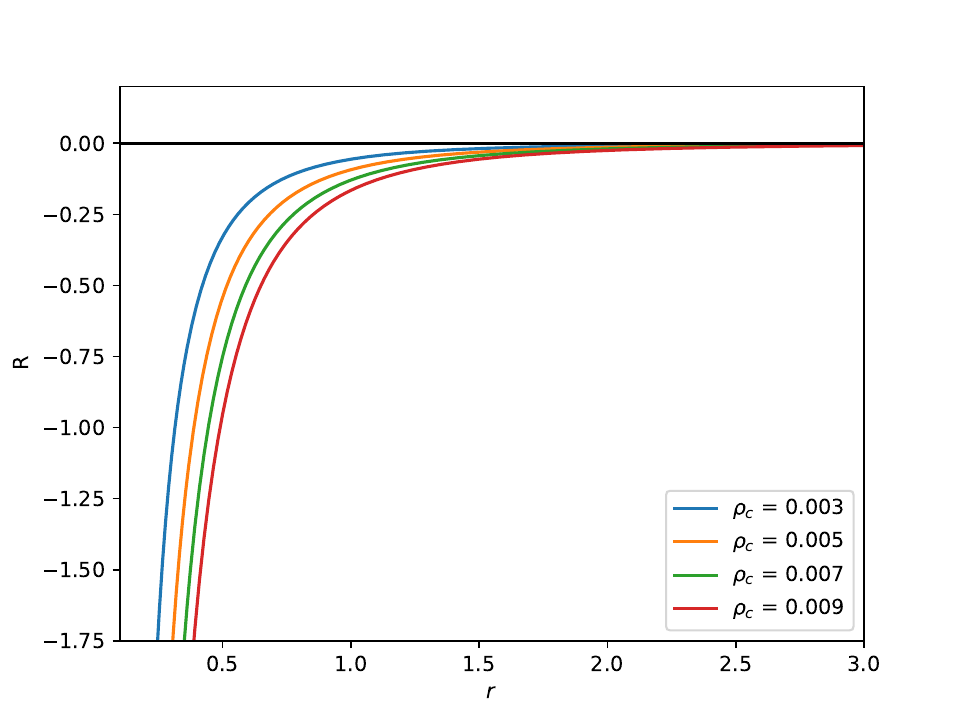}
\includegraphics[width=7.5cm]{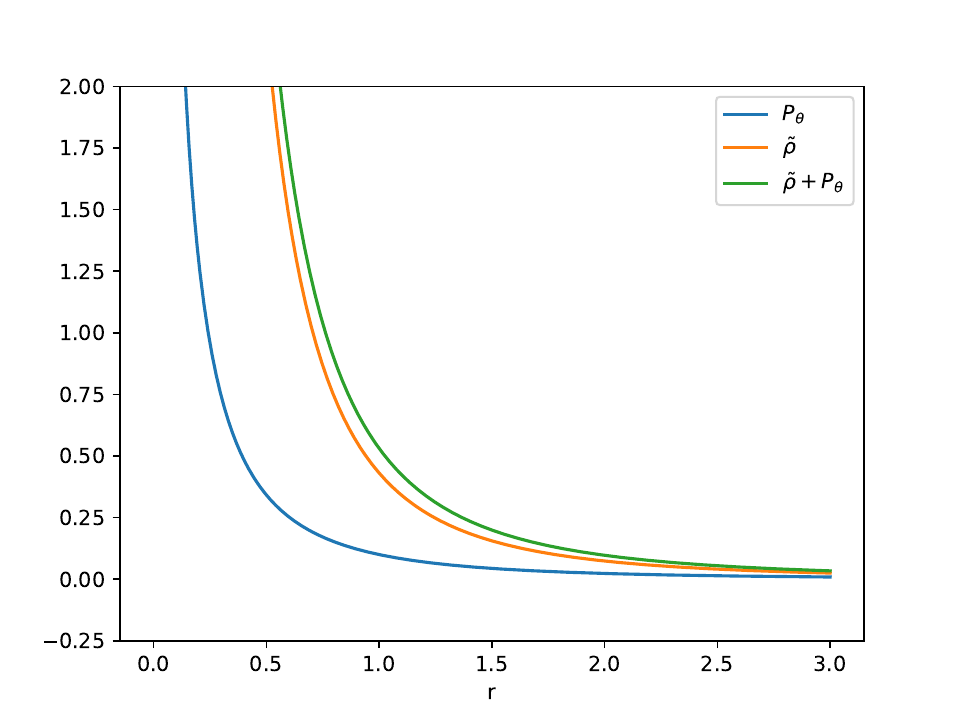}
        \caption{Left panel is the variation of Ricci scalar with respect to the radial coordinate 
	$r$ with various values of dark matter critical densities $\rho_c$ and other parameters 
	considered to be one. Right panel represents 
	the variation of Einstein's energy density and pressure.}
        \label{RicciS_vs_rh}
\end{figure}
%%%%%%%%

 From the Einstein tensor, we have calculated the density and pressure of the energy momentum tensor of rotating dark matter case as given below:
\begin{eqnarray}
        \rho &=& -P_r = -\frac{(a^2-r^2-\Delta(r)+r\partial_r \Delta(r)}{\Sigma(r,\theta)^4} \nonumber \\
        P_{\theta} &=& P_{\phi} = -\frac{(a^2(1+\cos^2\theta)-\Delta(r)+r\partial_r \Delta(r))}{\Sigma(r,\theta)^4} + \frac{\partial_r^2 \Delta(r)}{2 \Sigma(r,\theta)^2} \nonumber\\
        &=& -P_r +\frac{1}{2 \Sigma(r,\theta)^2} \big(\partial_r^2 \Delta(r) - 2\big)
        \label{density_pressure}
\end{eqnarray}
It is clear from the above expression eqn.(\ref{density_pressure}) that the physical ring
singularity is also present in the energy density and the pressure. In the right panel of
the figure \ref{RicciS_vs_rh}, we have plotted $P_{\theta}$ and $\rho + P_{\theta}$ and showed
that both are positive outside the horizon.

\subsection{Event horizon and Ergosphere}
Here we will discuss briefly the horizons and ergosphere region of the rotating black hole 
surrounded by the cold dark matter. They play crucial role (specially event horizon ($r_h^+$))
to find the thermodynamic variables of the black hole.

The condition to find the event horizon radius is
$g^{rr} = 0$ which gives
\begin{equation}
	\Delta(r) = 0.
\end{equation}
Event horizon radius can be calculated from 
the eqn.(\ref{delta_cdm}) as given below
\begin{eqnarray}
a^2 - 2 M r + r^2 \Big(1+{r\over R_{s}}\Big)^{-{8\pi \rho_c R^3_{s}\over r}} = 0.
	\label{horizon_soln}
\end{eqnarray}
%%%%%%%%%%
\begin{figure}
\centering
\includegraphics[width=7.5cm]{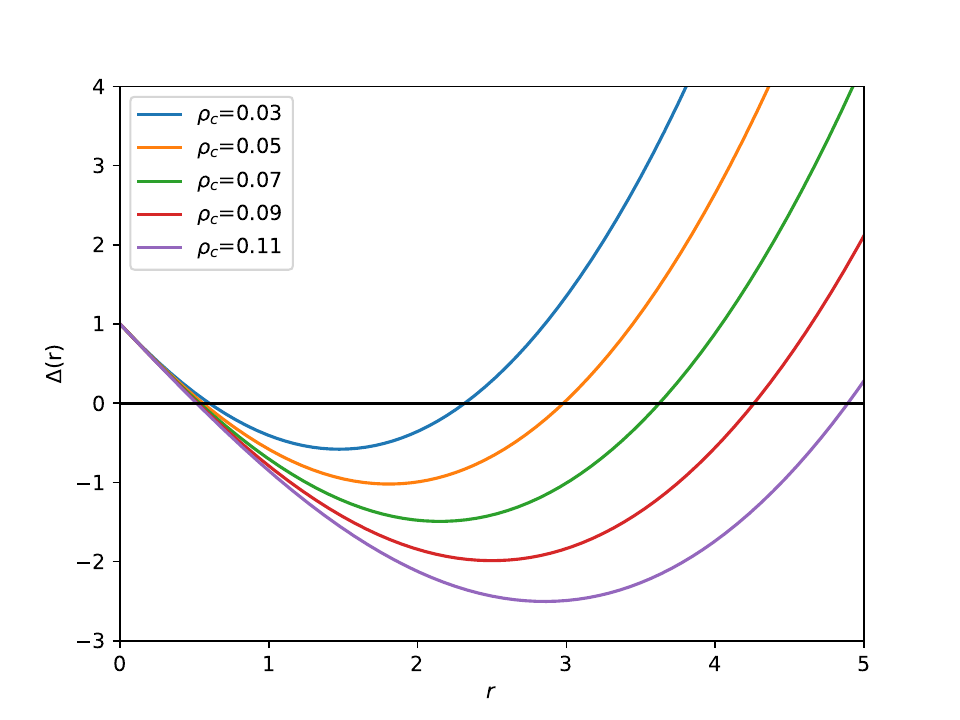}
	\caption {Variation of $\Delta(r)$ with respect to radial coordinate $r$ with various dark matter density. }
	\label{Delta_vs_rh}
\end{figure}
%%%%%%%%

The exact analytical solution of eqn.(\ref{horizon_soln}) can not be found but one can solve
numerically. However the horizon radius can be put in the simple form as given below 
\begin{equation}
	r_h = M \pm M\sqrt{1 - {a^2 - r_h^2 + r_h^2 \Big(1+{r_h\over R_{s}}\Big)^{-{8\pi \rho_c R^3_{s}\over r_h}}\over M^2}}
\end{equation}
In the figure(\ref{Delta_vs_rh}), we have plotted the $\Delta(r)$ with respect to radial coordinate
$r$ with various values of critical densities of cold dark matter and showed that there are two 
horizons one is called inner horizon ($ r_h^{-}$ cauchy horizon) and other one is called outer 
horizon ($r_h^+$, event horizon). In this figure, the points where change in sign indicate the 
horizon radius. Both the horizon depend nontrivially on the dark matter parameter ($\rho_c$). 
Horizon radii increase with increase of critical density ($\rho_c$) as shown in the figure \ref{Delta_vs_rh}.

\begin{figure}
\centering
\includegraphics[width=5cm]{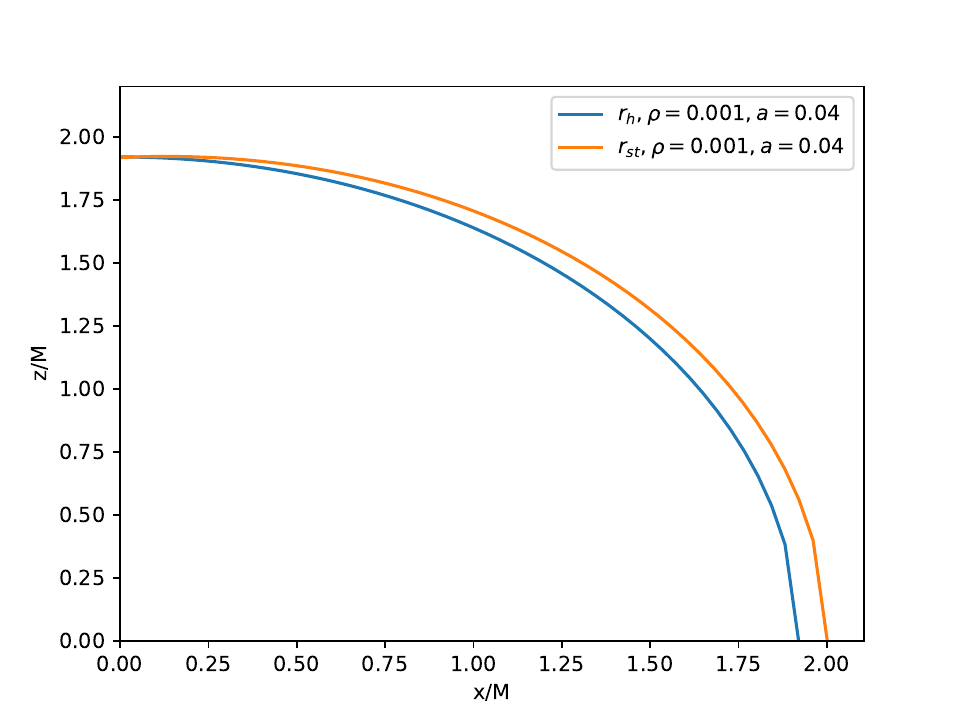}
\includegraphics[width=5cm]{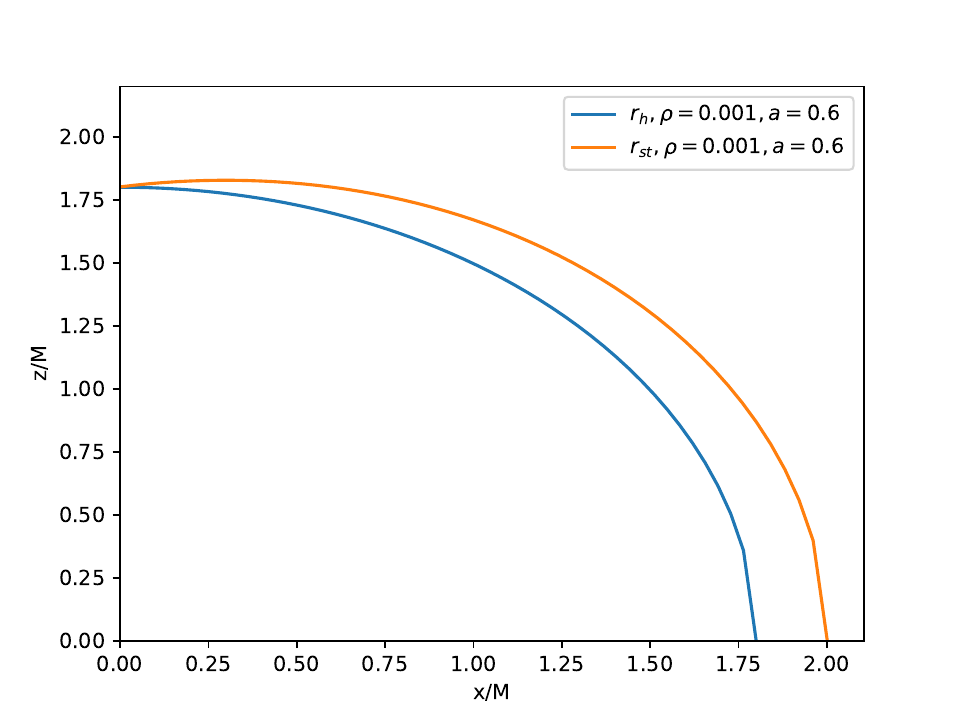}
\includegraphics[width=5cm]{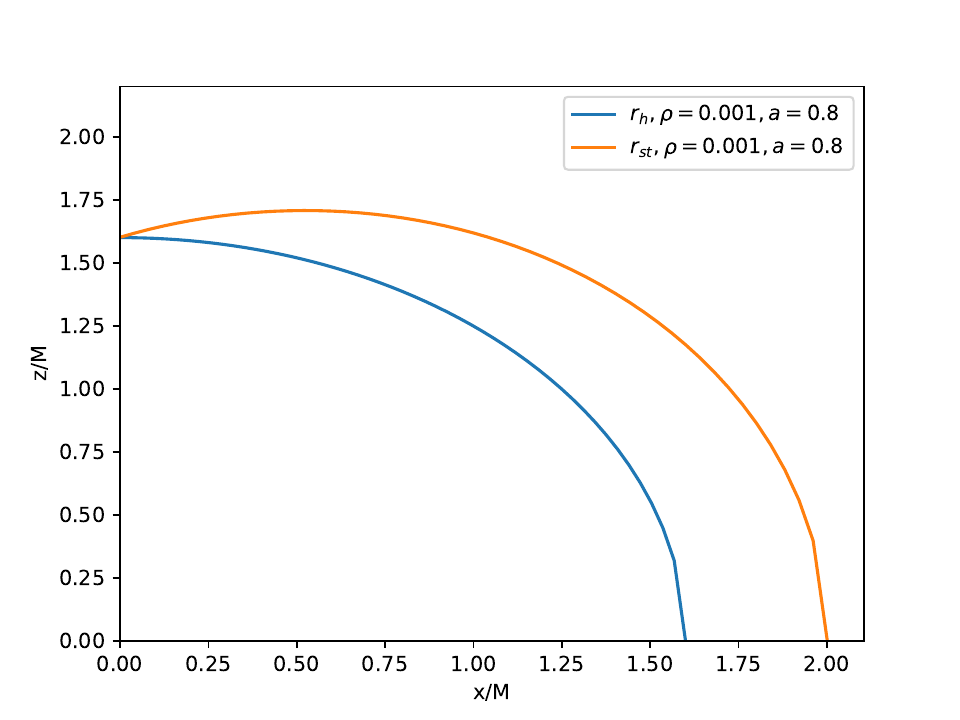}
\includegraphics[width=5cm]{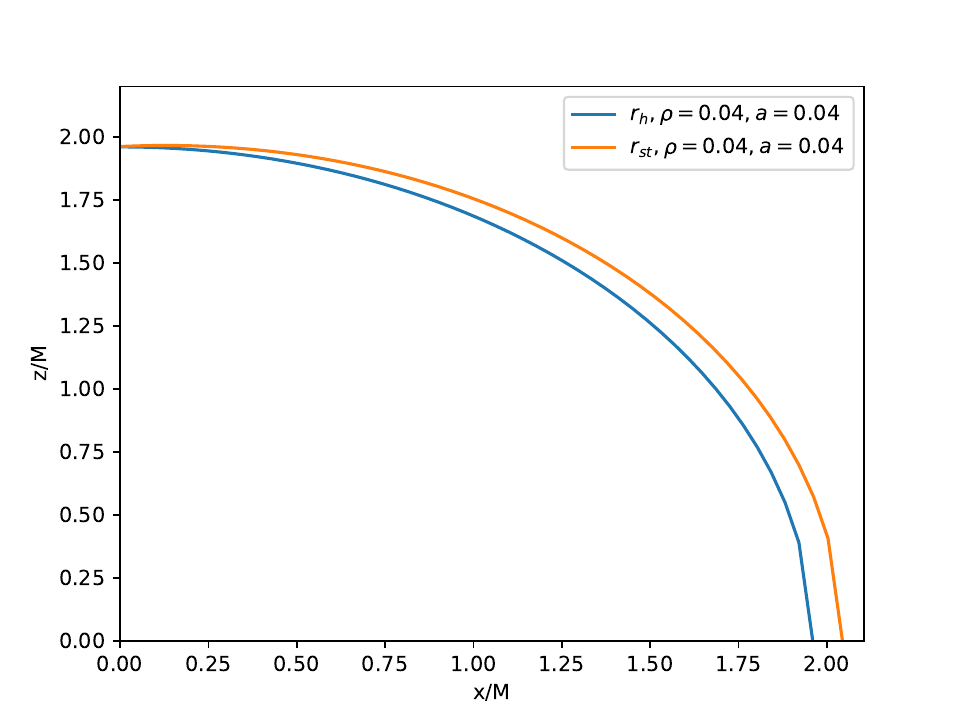}
\includegraphics[width=5cm]{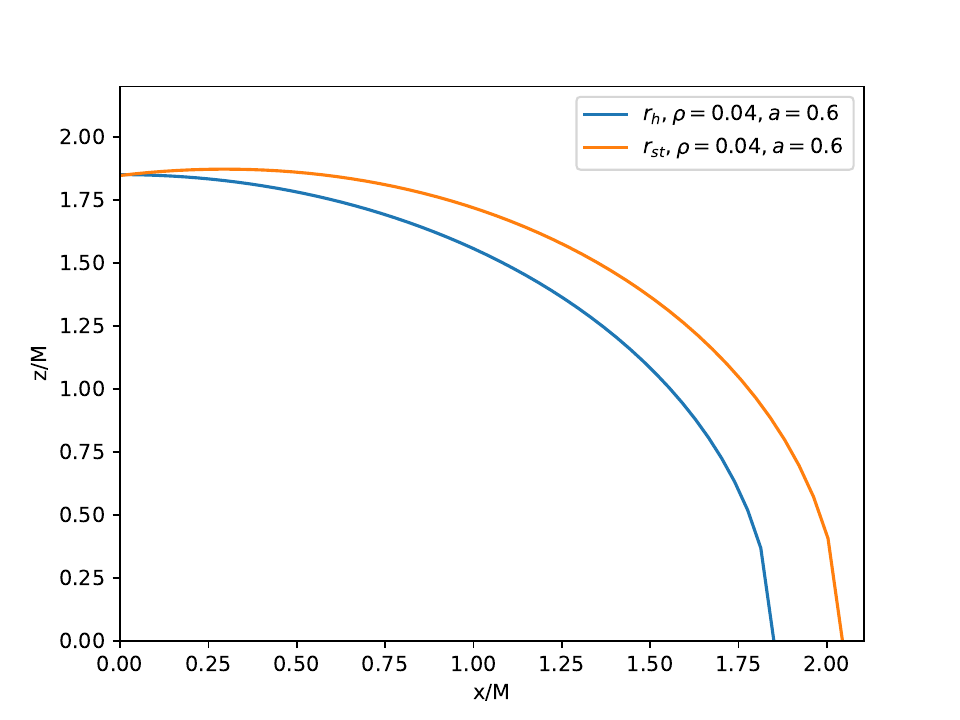}
\includegraphics[width=5cm]{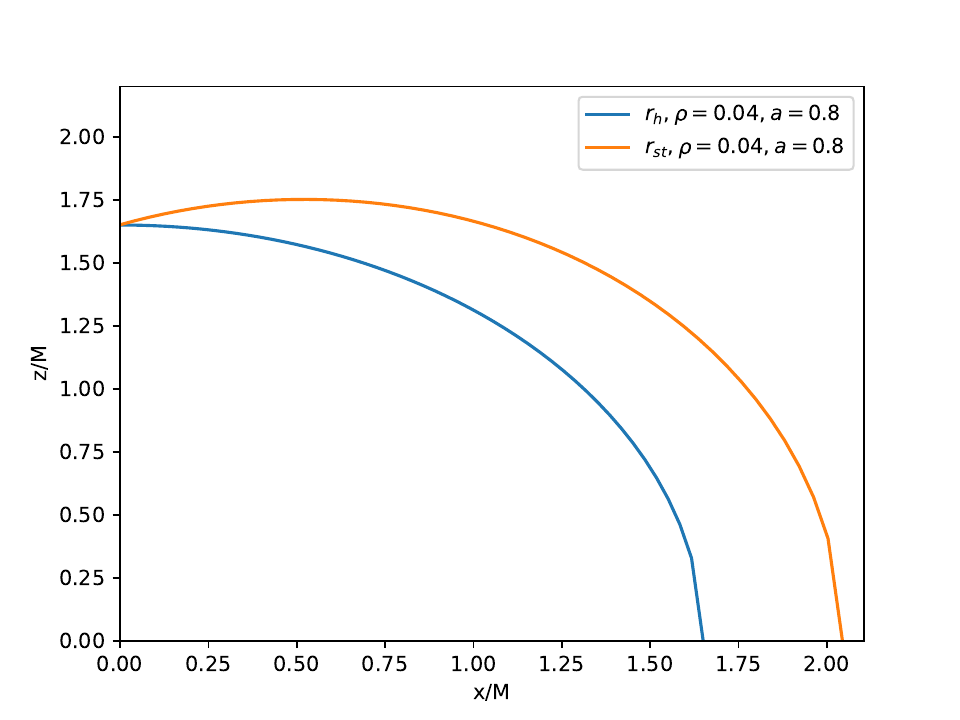}
\includegraphics[width=5cm]{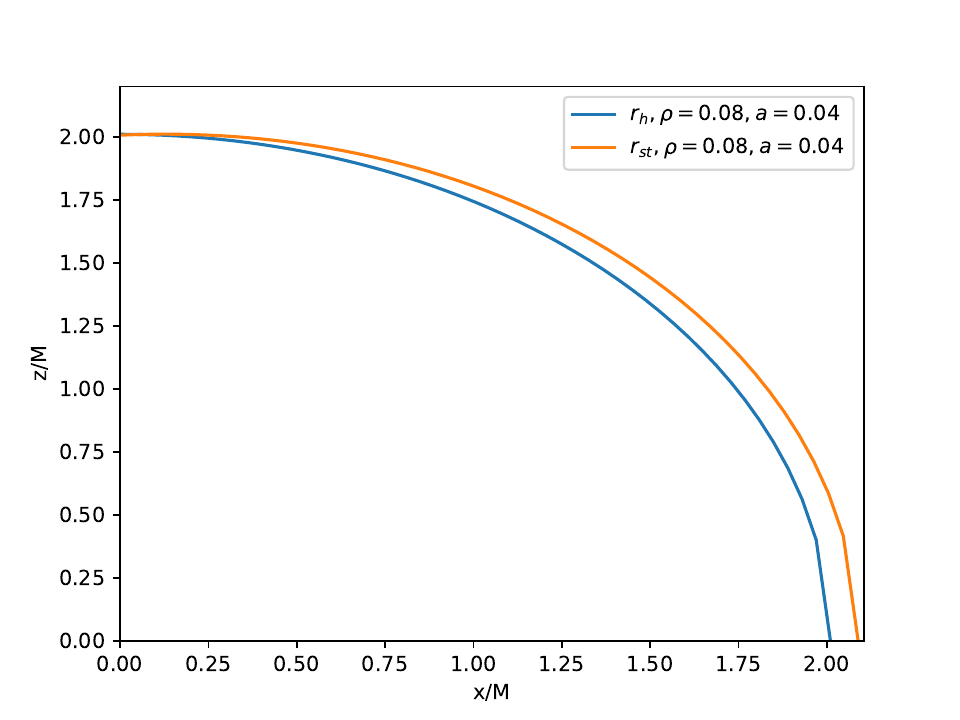}
\includegraphics[width=5cm]{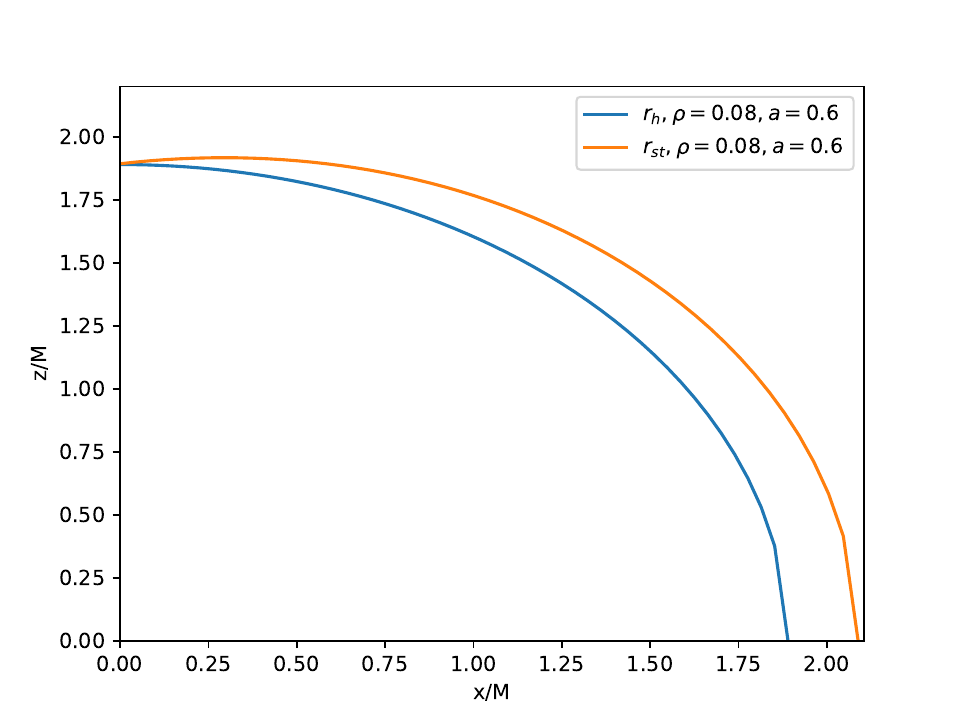}
\includegraphics[width=5cm]{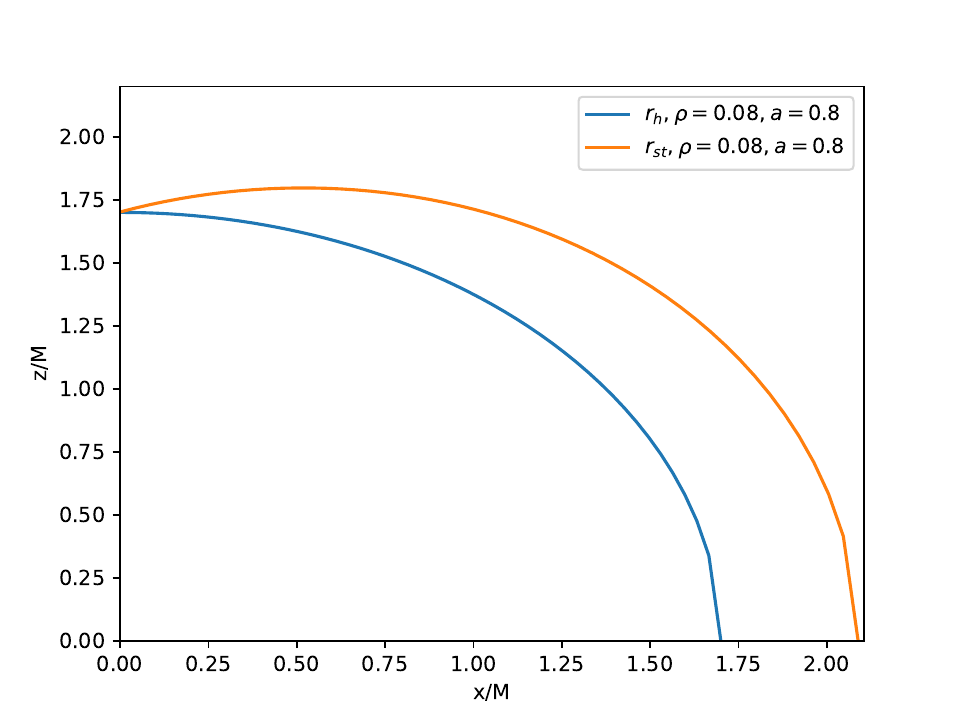}
	\caption{Behavior of ergoregion in the xz plan with different values of critical densities}
	\label{ergo_sphere}
\end{figure}

Ergosphere is a region between the event horizon and static limit surface. The static limit 
surface is defined by $g_{tt} = 0$ i.e.
\begin{equation}
	\Delta(r) - a^2\sin^2\theta = 0.
	\label{static_limit}
\end{equation}
Similar to the event horizon solution, static limit surface has two solutions ($r^-_{sls}, r^+_{sls}$).
The region between the event horizon  ($r^+_h$) and positive root of static limit surface 
($r^+_{sls}$) is the ergosphere region.
 Comparing eqns(\ref{horizon_soln},\ref{static_limit}), we find that both horizon radius and 
 static limit radius coincide each other at $\theta = 0\, \rm{or}\, \pi$.
 In the figure \ref{ergo_sphere}, we have plotted event horizon and ergosphere in the x-z plane 
 with various values of rotational parameter $a$ and critical density of cold dark matter. 
 At the equitorial plane ($\theta = \pi/2$), static radius does not depend on the rotational 
 parameter $a$. As we increase the value of rotation parameter $a$, the horizon radius ($r_h$) 
 decreases hence the region of  ergospace increases.
%%%%%%%%%%
%%%%%%%%

%%%%%%%%%%%%%%
{\section{Thermodynamics of rotating black hole in the presence of Dark Matter}
\label{thermodyn-computation}}
%%%%%%%%%%%%%%
In the section, we have discussed the thermodynamical properties of rotating black hole in the presence of cold dark matter. We have computed all the thermodynamics variables like Bekenstein-Hawking entropy, Hawking's temperature, Gibb's free enegry and Specific heat at the event horizon  ($\Delta(r)=0$).

For the given metric in eqn.(\ref{cdm_metric}), the Bekenstein-Hawking entropy of the black hole 
can be calculated from the area of the horizon as
\begin{equation}
	S_{BH} = {1\over 4 G_N} \int \, d\theta\, d\phi \sqrt{g_{\theta\theta}g_{\phi\phi}}
	= {1\over 4 G_N} \int \, d\theta\, d\phi\, \sin\theta \sqrt{(r^2+a^2)^2 -\Delta a^2\sin^2\theta}
\end{equation}
At the horizon $\Delta(r_h) = 0$, after integration one can get the simple form as given below
\begin{equation}
	S_{BH} = {\pi (r_h^2+a^2) \over G_N}
	\label{bh_entropy}
\end{equation}
From the above eqn(\ref{bh_entropy}), it apperently seems that the Bekenstein-Hawking 
entropy $S_{BH}$ does not depend on the any of the 
dark matter halo parameters like dark matter critical density, $\rho_c$ or scale radius of the 
dark matter halo $R_s$ explicitly. However, it does depend on the these parameters through 
horizon radius.

Now we want to calculate the hawking temperature of the rotating black hole in the presence of
cold dark matter.
The hawking temperature of the rotating black hole can be calculated various methods like by 
computing the surface gravity \cite{fatima,s.h.hendi} or by the standard 
Euclidean procedure \cite{g.w.gibbons,r.b.mann}. In this article, we have follow the the standard
Euclidean procedure. For completeness, we have briefly reviewed the procedure.
We defined two orthogonal killing vectors
\begin{eqnarray}
	\xi &=& \partial_t + {a\over a^2+r^2} \partial_{\phi}, ~~~ \tilde{\xi} = a\,\sin^2\theta\,\partial_t +\partial_{\phi} \nonumber \\[1ex]
	\xi^2 &=& -{\Delta\Sigma^2 \over (a^2+r^2)^2}, ~~~ \tilde{\xi}^2 = \Sigma^2\, sin^2\theta, ~~~ 
	\xi.\tilde{\xi} = 0
\end{eqnarray}
The vectors $\xi$ and $\tilde{\xi}$ are the dual to each other and  null on the horizon 
i.e. $\xi^2(r_h) = 0$, $\tilde{\xi}^2(r_h) = 0$. 
It is assume that $\xi$ is time-like everywhere outside the
horizon $r \ge r_h$ , whereas the vector $\tilde{\xi}$ is space-like everywhere outside the axis
$(\theta = 0,~\theta = \pi)$ where $\tilde{\xi}^2 = 0$. The {\it 1-form} corresponding
to $\xi$ and $\tilde{\xi}$ are
\begin{eqnarray}
	d\omega = {a^2+r^2\over \Sigma^2}(dt - a\sin^2\theta\, d\phi), ~~~ 
	d\tilde{\omega} = {a^2+r^2\over \Sigma^2}(-{a\over a^2+r^2}dt +  d\phi)
\end{eqnarray}
With standard Euclideanization of time variable $t = i \tau$ and along with the transformation 
of the rotating parameter $a = i\hat{a}$, the Euclidean killing vectors $\xi, \tilde{\xi}$ and their dual {\it 1-form} $d\omega, \tilde{d\omega}$ take form
\begin{eqnarray}
	\xi &=& \partial_{\tau} + {\hat{a}\over \hat{a}^2+r^2} \partial_{\phi}, ~~~ \tilde{\xi} = \hat{a}\,\sin^2\theta\,\partial_{\tau} +\partial_{\phi} \nonumber \\[1ex]
	d\omega &=& {r^2-\hat{a}^2\over \hat{\Sigma}^2}(d\tau - \hat{a}\sin^2\theta\, d\phi), ~~~
	d\tilde{\omega} = {r^2-\hat{a}^2\over \hat{\Sigma}^2}({\hat{a}\over r^2-\hat{a}^2}d\tau +  d\phi)
\end{eqnarray}
where $\hat{\Sigma}^2 = r^2 - \hat{a}^2 \cos^2\theta$. The Euclidean metric reduces to 
\begin{equation}
	ds_E^2 = - {\hat{\Delta}\hat{\Sigma^2} \over (r^2-\hat{a}^2)^2}d\omega^2 - {\hat{\Sigma^2}\over \hat{\Delta}} dr^2 - \hat{\Sigma^2} (d\theta^2+\sin^2\theta\,d\tilde{\omega})
\end{equation}
with $\hat{\Delta} = -\hat{a}^2 - 2 M r + r^2 \Big(1+{r\over R_s}\Big)^{-{8\pi \rho_c R^3_s\over r}}$
which will give the horizon radius $r = r_h^{+}(\hat{r}_h) $. Introducing new radial cooridnate $x$ near the horizon
\begin{eqnarray}
	\hat{\Delta} &=& \gamma(r-\hat{r}_h) = {\gamma^2 x^2 \over 4}\nonumber \\[1ex]
	\gamma &=& (\hat{r}_h-r_{-})  
\end{eqnarray}
the Euclidean metric near horizon becomes
\begin{equation}
	ds_E^2 = -\hat{\Sigma}^2_h \Big( dx^2 + {\gamma^2 x^2 \over 4 (\hat{r}_h^2-\hat{a}^2)^2} d\omega^2 \Big) - \hat{\Sigma}^2_h (d\theta^2 + \sin^2\theta \, d\tilde{\omega}^2)
\end{equation}
%where $\hat{\Sigma}^2 = \hat{r}_h^2 - \hat{a}^2 \cos^2\theta$. 
On the horizon surface ${\cal H}$, 
introducing a new coordinate $\psi = \phi + {\hat{a}\tau \over \hat{r}^2_h-\hat{a}^2}$, 
the surface horizon metric can be written as 
\begin{equation}
	ds^2_{\cal H} = \hat{\Sigma}^2_h (d\theta^2 + \sin^2\theta \, d\tilde{\omega}^2) = \hat{\Sigma}^2_h (d\theta^2 + {(\hat{r}_h-\hat{a}^2)^2 \over \hat{\Sigma}^4_h} \sin^2\theta \,d\psi)
\end{equation}

The points $\psi$ and $\psi + 2\pi$ should be identified on the horizon ${\cal H}$ for the regularity of the metric at points $\theta = 0$ and $\theta = \pi$. The full metric now turn into
\begin{equation}
	ds^2_E = -\hat{\Sigma}^2_h\,ds^2_{C_2} - ds^2_{\cal H}
\end{equation}
where $ds^2_{C_2}$ is the metric on the two dimensional disk $C_2$ given by
\begin{equation}
ds^2_{C_2} = dx^2 + {\gamma^2 x^2 \over 4 (\hat{r}_h^2-\hat{a}^2)^2} d\omega^2
\end{equation}
%Considering the above metric with fixed $(\theta, \phi)$ and 
By introducing a new angle coordinate
$\chi = \tau - \hat{a}\sin^2\theta\, \phi$ the metric on the disk can be put in the following form
\begin{equation}
	ds^2_{C_2} = dx^2 + {\gamma^2 x^2 \over 4 \hat{\Sigma}^4_h} d\chi^2 = dx^2 + x^2 d\Big({\gamma \over 2 \hat{\Sigma}^2_h} \chi\Big)^2
\end{equation}
This metric is very similar form as plane polar coordinate if the angle variable, ${\gamma \over 2 \hat{\Sigma}^2_h}$ 
has the period of $2\pi$ and has a conical singularity at $x=0$. To avoid this conical singularity
one needs to 
identify the points $\chi$ and $\chi + 4\pi\hat{\Sigma}^2_h \gamma^{-1}$.
%%%%%%%%%%
\begin{figure}
\centering
\includegraphics[width=7cm]{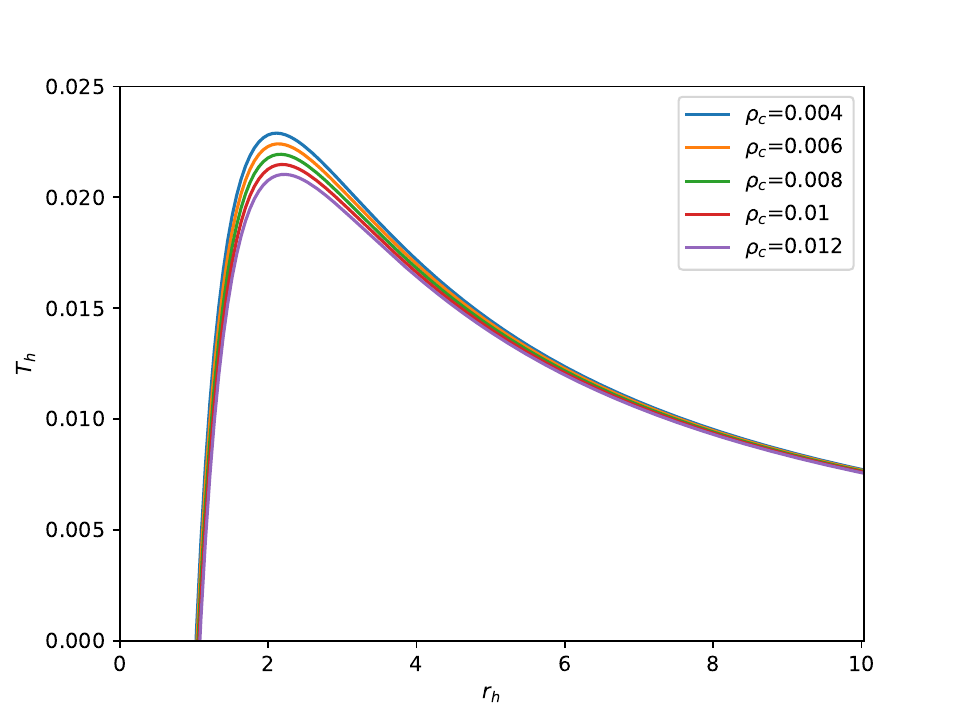}
	\caption{variation of Hawking temperature ($T_h$) with respect to 
	horizon radius with variour values of cold dark matter critical density ($\rho_c$).}
	\label{Th_rh}
\end{figure}
%%%%%%%%
%This must be true independently of the coordinate $\theta$ on the horizon ${\cal H}$, one must also
This turns out to be
the points $(\tau, \phi)$ and $(\tau - \beta_h,\, \phi - \Omega_h\, \beta_h)$ must be identified
on the horizon ${\cal H}$, where 
$\beta_h = 4\pi\hat{\Sigma}^2_h \gamma^{-1}$ is the inverse hawking temperature of the black hole 
and $\Omega = \hat{a}/(\hat{r}^2-\hat{a}^2)$ is the angular velocity. By analytically continued
to the real values $a$, the hawking temperature 
of rotating black hole with cold dark matter is given as
\begin{eqnarray}
	T_h &=& {1\over \beta_h} = {1\over 4\pi}{\gamma \over r_h^2 + a^2} %\nonumber \\[1ex]
	%\Omega_h &=& {a\over r_h^2+a^2}.
\end{eqnarray}
where 
\begin{equation}
	\gamma = {1\over r_h} \Bigg[\Big(1+r_h/R_s\Big)^{-8\pi\rho_c R_s^3 \over r_h} \Bigg(r_h^2 + 8\pi\rho_c R_s^3 \Big(r_h\ln(1+r_h/R_s) - {r_h^2\over  r_h+R_s}\Big) \Bigg) - a^2 \Bigg]
\end{equation}

 In figure\ref{Th_rh}, we have plotted Hawking temperature with respect to horizon radius for 
 various values of critical density of CDM. There are two black hole branches at a given 
 temperature. The first branch, where the temperature increases 
 with radius, is stable and occurs for small values of $r_h$. Conversely, the second branch, 
 where the temperature decreases with radius, is unstable and occurs for large $r_h$. 
 The thermodynamic stability or instability of these small and 
 large black hole branches will be further confirmed through an analysis of 
 free energy and specific heat.

The Gibbs free energy is given by\cite{s.mahapatra,dvsingh}: 
\begin{equation}
 {\cal G} = {\cal M} - T_h S_{BH}
\end{equation}
%%%%%%%%%%
\begin{figure}
\centering
\includegraphics[width=7cm]{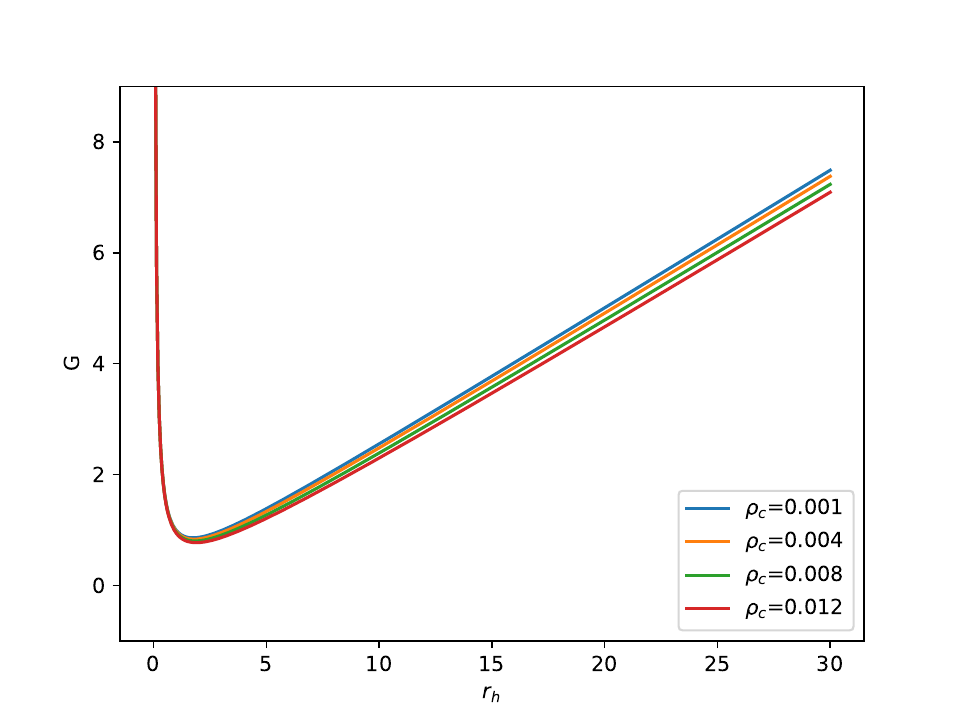}
\includegraphics[width=7cm]{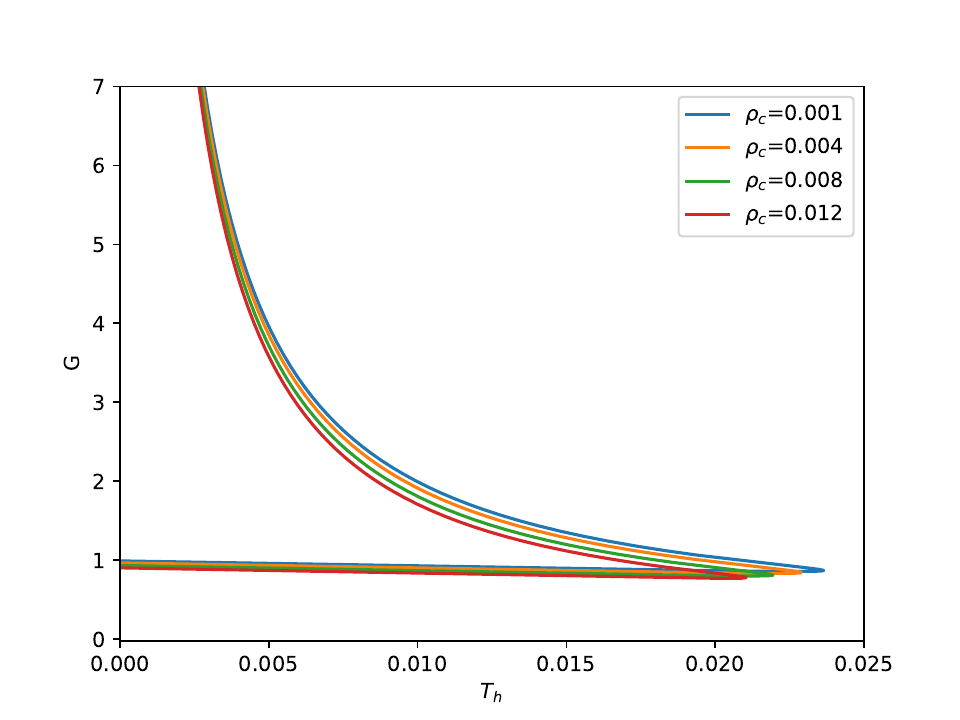}
	\caption{Left panel is the variation of Gibb's free energy (${\cal G}$) with respect to 
	horizon radius with variour values of cold dark matter critical density ($\rho_c$) and 
	right panel represents the variation of Gibb's free energy (${\cal G}$) with respect to
	Hawking temperature ($T_h$).}
	\label{GvsTh}
\end{figure}
%%%%%%%%
 In figure(\ref{GvsTh}), we have plotted Gibb's free energy ({\cal G}) with respect to the 
 Hawking temperature ($T_h$) with different values of critical densities ($\rho_c$) of cold 
 dark matter. In this figure, the horizon radius $r_h$ increases from left to right.
 For a fixed temperature, there are two black hole branches. The lower branch corresponds 
 to more stable black hole with smaller Gibb's free energy and positive specific heat ($C_p>0$). 
 Whereas the upper branch corresponds to unstable black hole (Schwarzschild-like black hole) 
 with higher free energy and negative specific heat ($C_p<0$).

%%%%%%%%%%
\begin{figure}
\centering
\includegraphics[width=7cm]{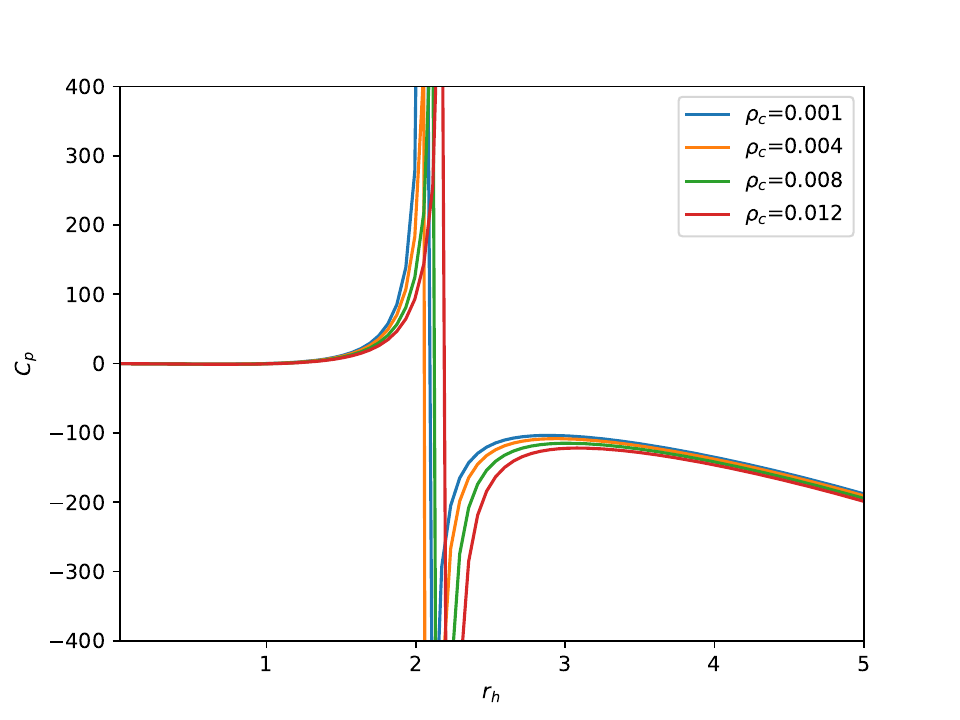}
	\caption{Variation of specific heat ($C_p$) with 
	respect to horizon radius ($r_h$).}
	\label{Cp_vs_rh}
\end{figure}
%%%%%%%%
In figure(\ref{Cp_vs_rh}), we have plotted specific heat ($C_p = T_h (\partial S_{BH} / \partial T_h)$) with respect to the horizon radius 
with various values of critical densities ($\rho_c$) of cold dark matter. From this figure we 
notice that specific heat is positive for small black hole branch and negative for large black hole branch as we have stated earlier. So positivity of specific heat in small black hole branch shows thermodynamically stable in nature whereas negativity of specific heat in large black hole branch implies the thermodynamically stable in nature.

\section{Conclusions}

  In this article, we have computed geometrical quantities like Ricci scalar Kretschmann scalar, 
  Einstein energy momentum tensor of a rotating black hole in the presence of cold dark matter and 
  showed that it is very similar to the $Kerr's$ rotating black hole in the critical 
  limit ($\rho_c \rightarrow 0$). We have studied the effect of cold dark matter critical density
  on event horizons and ergosphere region. We find that by increasing the critical density, 
  increases the border of both the horizons and ergosphere for a fixed value of rotational 
  parameter $a$. It is also noticed that for a fixed value of critical density ($\rho_c$), 
  the horizon radius decreases with increasing the rotational parameter $a$ and hence the area 
  between the horizon and static limit surface inceases. 

We also studied the thermodynamic stability of rotating black hole with cold dark matter scenario.
The qualitative features remain same as $Kerr's$ case but quantitatively found corrections due to 
cold dark matter presence. 
Specifically, we observe thermodynamically stable small black hole branches alongside unstable 
large ones similar to the $Kerr's$ case. This feature is clearly visible in all the thermodynamical 
variables, noticiably in Gibb's free energy (${\cal G}$) and Specific heat 
($C_p$). We have also discovered that the stability of the rotating black hole in the presence of
 cold dark matter, within certain horizon radius and temperature ranges, is intricately 
 influenced by the parameter critical density. 

   There are numerous intriguing avenues to explore further in this study. It would be intersting 
   to study the dynamical stability of these black holes against various perturbations. 
   Another intriguing area to explore involves investigating thermal fluctuations,
   by incorporating the system's corrected entropy using partition function\cite{m.sharif,b.pourhassan,s.gunasekaran}.
   It would be compelling to examine how the presence of the cold dark matter impacts various 
   phenomena such as the black hole shadow, gravitational waves, photon orbits, 
   quasi-normal modes, and Hawking radiation, and to discern these effects from those 
   observed in the Kerr case. We are considering these issues for our future investigation.

\section*{Acknowledgements}
We would like to thank Dr. Anindya Biswas for useful discussion.
SM would like to thank the Department of Physical Sciences, IISER Kolkata 
for hospitality while part of the project was being carried out.
SM also acknowledges the Science and Engineering research Board, India for 
financial assistance under the Teachers Associateship for Research Excellence (TARE) grant.

%%%%%%%%%%%%%%%%%%%%%%%%%%%%%%%%%%%%%%%%%%%%%%%%%%%%%%%%%%%%%%%%%%%%%%%%
%                                                                      %
%       %%%     Define Journal macros                                  %
% \newcommand{\araa}[3]{{\em Annu. Rev. Astron. Astrophys.\/}
%          {\bf#1} (19#3) #2}                                          %
% \newcommand{\ptp}[3]{{\em Prog. Theoret. Phys. (Kyoto)\/}
%          {\if#1} (19#3) #2}                                          %
\newcommand{\plb}[3]{{Phys. Lett.} {\bf B#1} (#3) #2}                  %
\newcommand{\prl}[3]{Phys. Rev. Lett. {\bf #1} (#3) #2}        %
\newcommand{\rmp}[3]{Rev. Mod.  Phys. {\bf #1} (#3) #2}             %
\newcommand{\prep}[3]{Phys. Rep. {\bf #1} (#3) #2}                     %
\newcommand{\rpp}[3]{Rep. Prog. Phys. {\bf #1} (#3) #2}             %
\newcommand{\prd}[3]{{Phys. Rev.}{\bf D#1} (#3) #2}                    %
\newcommand{\np}[3]{Nucl. Phys. {\bf B#1} (#3) #2}                     %
\newcommand{\npbps}[3]{Nucl. Phys. B (Proc. Suppl.)
           {\bf #1} (#3) #2}                                           %
\newcommand{\sci}[3]{Science {\bf #1} (#3) #2}                 %
\newcommand{\zp}[3]{Z.~Phys. C{\bf#1} (#3) #2}                 %
\newcommand{\mpla}[3]{Mod. Phys. Lett. {\bf A#1} (#3) #2}             %
\newcommand{\astropp}[3]{Astropart. Phys. {\bf #1} (#3) #2}            %
\newcommand{\ib}[3]{{\em ibid.\/} {\bf #1} (#3) #2}                    %
\newcommand{\nat}[3]{Nature (London) {\bf #1} (#3) #2}         %
\newcommand{\nuovocim}[3]{Nuovo Cim. {\bf #1} (#3) #2}         %
\newcommand{\yadfiz}[4]{Yad. Fiz. {\bf #1} (#3) #2 [English            %
        transl.: Sov. J. Nucl.  Phys. {\bf #1} #3 (#4)]}               %
\newcommand{\philt}[3]{Phil. Trans. Roy. Soc. London A {\bf #1} #2
        (#3)}                                                          %
\newcommand{\hepph}[1]{(electronic archive:     hep--ph/#1)}           %
\newcommand{\hepex}[1]{(electronic archive:     hep--ex/#1)}           %
\newcommand{\astro}[1]{(electronic archive:     astro--ph/#1)}         %

\newcommand{\jcap}[3]{Journal of Cosmology and Astroparticle Physics {#1}. 
  doi:10.1088/1475-7516/#1/#2/#3}                                          %
\newcommand{\apj}[3]{The Astrophysical Journal {#1}, {#2}. doi:10.1086/{#3}} %
\newcommand{\cqg}[3]{Classical and Quantum Gravity {#1}, (#3). {#2}}
\newcommand{\cqgr}[4]{Classical and Quantum Gravity {#1}, {#2}. doi:10.1088/0264-9381/{#1}/{#3}/{#4}}
\newcommand{\cmp}[3]{Commun. Math. Phys {#1:}(#3){#2}}
\newcommand{\xgray}[3]{X- and Gamma-Ray Astronomy {#1}(#3){#2}}
\newcommand{\araa}[3]{ Annual Review of Astronomy and Astrophysics {#1}(#3){#2}}
\newcommand{\ijmpd}[3]{ International Journal of Modern Physics {\bf D#1}(#3){#2}}
\newcommand{\ctp}[3]{ Communications in Theoretical Physics {\bf #1}(#3){#2}}
\newcommand{\epj}[3]{Eur.Phys.J. {\bf #1} (#3) {#2}}
\newcommand{\ijmp}[3]{ Int. J. Mod. Phys.{\bf #1}(#3){#2}}
%       \relax                                                         %
%       %%%     End     Journal macro definitions                      %
%                                                                      %
%%%%%%%%%%%%%%%%%%%%%%%%%%%%%%%%%%%%%%%%%%%%%%%%%%%%%%%%%%%%%%%%%%%%%%%%

\end{document}